\documentclass[aps,prl,twocolumn,superscriptaddress]{revtex4-2}
\usepackage{amsmath}
\usepackage{physics}
\usepackage{amsfonts}
\usepackage{bm}
\usepackage{graphicx}
\usepackage{hyperref}
\usepackage{pifont}
\usepackage{color}
\usepackage{tabularx}

\begin{document}

\author{Fengyuan Xuan}
\affiliation{Centre for Advanced 2D Materials, National University of Singapore, 6 Science Drive 2, Singapore 117546}
\affiliation{Suzhou Laboratory, Suzhou, 215123, China}

\author{MingRui Lai}
\affiliation{Integrative Sciences and Engineering Programme, NUS Graduate School, National University of Singapore, Singapore 119077}

\author{Yaze Wu}
\altaffiliation[Current affiliation: ]{Institute of High Performance Computing (IHPC), Agency for Science, Technology and Research (A$^*$STAR), 1 Fusionopolis Way, \#16-16 Connexis, Singapore 138632, Republic of Singapore}
\affiliation{Department of Physics, National University of Singapore, 2 Science Drive 3, Singapore 117551}
\affiliation{Centre for Advanced 2D Materials, National University of Singapore, 6 Science Drive 2, Singapore 117546}

\author{Su Ying Quek}
\email[Corresponding author: ]{phyqsy@nus.edu.sg}
\affiliation{Department of Physics, National University of Singapore, 2 Science Drive 3, Singapore 117551}
\affiliation{Centre for Advanced 2D Materials, National University of Singapore, 6 Science Drive 2, Singapore 117546}
\affiliation{Integrative Sciences and Engineering Programme, NUS Graduate School, National University of Singapore, Singapore 119077}
\affiliation{Department of Materials Science and Engineering, National University of Singapore, Singapore 117575}

\title{Exciton-Enhanced Spontaneous Parametric Down-Conversion in Two-Dimensional Crystals}

\date{\today}

\begin{abstract}
We show that excitonic resonances and interexciton transitions can enhance the probability of spontaneous parametric down-conversion, a second-order optical response which generates entangled photon pairs.
We benchmark our \textit{ab initio} many-body calculations using experimental polar plots of second harmonic generation in NbOI$_2$, clearly demonstrating the relevance of excitons in the nonlinear response.
A strong double-exciton resonance in 2D NbOCl$_2$ leads to giant enhancement in the second order susceptibility.
Our work paves the way for the realization of efficient ultrathin quantum light sources. 
\end{abstract}

\maketitle

Spontaneous parametric down-conversion (SPDC) is a second-order nonlinear optical process in which a pump photon with frequency $\omega_p$ spontaneously splits into two correlated photons - a signal photon with frequency $\omega_s$ and an idler photon with frequency $\omega_i=\omega_p-\omega_s$ ~\cite{dorfman2016nonlinear,reverseCouteau,pan2012multiphoton}. Such entangled photons are fundamental to the realization of quantum communication and quantum information applications~\cite{pan2012multiphoton}.
The signal and \textcolor{black}{idler} frequencies can be the same (degenerate SPDC), or different (non-degenerate SPDC). 
Just recently, entangled photon pairs were generated through SPDC in 46-nm-thick NbOCl$_2$~\cite{guo2023ultrathin}.  
Traditionally, bulk materials are used for the generation of entangled photon pairs, because the probability amplitude for this process scales linearly with volume. 
Moving to the 2D limit is, however, at the forefront of experimental research in nonlinear optics, due to inherent advantages such as reduction in losses due to optical absorption, and more facile integration with hybrid quantum photonic platforms which can enhance the incident light intensities~\cite{ResNanoPh2021,ResOpt2021,ResACSnano2022}. 
Importantly, the relaxation of phase-matching conditions in 2D is not only helpful from a technical viewpoint, but also broadens the frequency spectrum for entangled photon pairs~\cite{PRLthin}. 

The theory of the general nonlinear optical response in semiconductors has been studied mostly within the independent particle approximation (IPA)~\cite{levine1990one,sipe1993nonlinear,luppi2010ab,young2012first,rangel2017large,Thygesen2021ACSnano}.
However, in layered 2D materials, dielectric screening is reduced, and electron-hole (or excitonic) interactions dominate the linear optical response~\cite{wang2018colloquium}. 
It is therefore critical to understand the effect of excitons on the nonlinear optical response. 
A number of studies on nonlinear optical properties \cite{attaccalite2011real,attaccalite2013nonlinear,kolos2021giant,chan2021giant} have been performed using real-time approaches that account for excitonic effects. However, these real-time approaches do not provide clear insights into the origin of the peaks in the nonlinear susceptibilities. 
On the other hand, spectra for second-harmonic-generation (SHG), the inverse process of degenerate SPDC, were computed for bulk zinc-blende structures within a first principles $GW$ plus Bethe-Salpeter-Equation ($GW$-BSE) approach in the frequency domain~\cite{2005PRB,chang2001excitonic,riefer2017solving}, but the origin of the SHG peaks was not discussed. Some excitonic effects were observed in the SHG spectra, albeit not drastic~\cite{2005PRB,chang2001excitonic,riefer2017solving}.

In this work, we evaluate the Lehmann representation of the second-order susceptibility, using as a basis the many-body excitations obtained from $GW$-BSE.
Applying the approach to monolayer NbOX$_2$ (X = I, Cl), prototypical 2D materials of significant current interest in nonlinear optics~\cite{abdelwahab2022giant,abdelwahab2023highly,guo2023ultrathin,AdvMaterNbOI2},
we show that the probability amplitude of generating entangled photon pairs is significantly enhanced by excitonic effects. 
Interexciton transitions together with excitonic resonances lead to order-of-magnitude enhancements in the second-order optical response. 
We benchmark our approach by comparing our computed polarization-dependent SHG intensities with recent experiments~\cite{abdelwahab2022giant}; good agreement is obtained and the importance of excitons is clearly illustrated.
We further predict the existence of a double-resonance condition for NbOCl$_2$, that can dramatically increase the expected yield of entangled photon pairs compared to the non-resonant case.
This work illustrates the fundamental excitonic origins of SPDC in 2D layered materials, and demonstrates the potential of exciton-enhanced SPDC in 2D crystals, thus paving the way for further advancements in the field of 2D quantum photonics. 

Central to the theoretical description of SPDC is the second-order susceptibility, $\chi_{\alpha\beta\gamma}^{(2)}(\omega_3;\omega_1,\omega_2)$, arising from light-matter interaction~\cite{Mandel1985,pan2012multiphoton}. 
The Lehmann representation of $\chi_{\alpha\beta\gamma}^{(2)}(\omega_3;\omega_1,\omega_2)$ can be expressed as (summing over spin and setting $\hbar = 1$):
\begin{widetext}
\begin{equation}
\begin{split}
\chi_{\alpha\beta\gamma}^{(2)}(\omega_3;\omega_1,\omega_2) &= \frac{-ie^3}{m^3\Tilde{\omega}_3\Tilde{\omega}_1\Tilde{\omega}_2}\sum_{SS'}
\Bigg[\frac{\alpha_{0S} \beta_{SS'} \gamma_{S'0}}{(\Tilde{\omega}_2-\Omega_{S'})(\Tilde{\omega}_3-\Omega_S)} + \frac{\beta_{0S} \gamma_{SS'} \alpha_{S'0}}{(\Tilde{\omega}_1+\Omega_S)(\Tilde{\omega}_3+\Omega_{S'})} - \frac{\gamma_{0S} \alpha_{SS'} \beta_{S'0}}{(\Tilde{\omega}_1-\Omega_{S'})(\Tilde{\omega}_2+\Omega_{S})} \\
+&\frac{\alpha_{0S} \gamma_{SS'} \beta_{S'0}}{(\Tilde{\omega}_1-\Omega_{S'})(\Tilde{\omega}_3-\Omega_S)} + \frac{\gamma_{0S} \beta_{SS'} \alpha_{S'0}}{(\Tilde{\omega}_2+\Omega_S)(\Tilde{\omega}_3+\Omega_{S'})} - \frac{\beta_{0S} \alpha_{SS'} \gamma_{S'0}}{(\Tilde{\omega}_2-\Omega_{S'})(\Tilde{\omega}_1+\Omega_{S})}\Bigg]
\end{split}
\label{chi}
\end{equation}
\end{widetext}
where $\alpha_{SS'}$ refer to momentum matrix elements $\mel{S}{P^{\alpha}}{S'}$ and similarly for $\beta_{SS'}$ and $\gamma_{SS'}$. 
Here, $\Tilde{\omega}_1=\omega_1+i\eta$, $\Tilde{\omega}_2=\omega_2+i\eta$ and $\Tilde{\omega}_3=\Tilde{\omega_1}+\Tilde{\omega}_2$, 
and $\mathbf{P}$ is the many-body momentum operator, i.e. $\mathbf{P}=\sum_{i}\mathbf{p}_i$ where $\mathbf{p}_i$ is the single-particle momentum operator acting on particle $i$.
$\eta$ is a small positive number that describes the adiabatic switching-on of the electromagnetic field, and is bounded below by the inverse of the thermal equilibration time in the material~\cite{Giovanni2021}. 
This expression is obtained through second order perturbation theory~\cite{BOYD2008135,2005PRB}, which is valid for typical laser intensities~\cite{ahn2020low}.

The excitonic effect in $\chi^{(2)}$ is taken into account by defining $\Omega_S$ and $\ket{S}$ to be the many-body excitation energies and excited states obtained from $GW$-BSE within the Tamm-Dancoff approximation:
\begin{equation}
(E_{ck} - E_{vk})A^{S}_{vc\vb{k}} + \sum_{v'c'\vb{k}'}K_{vc\vb{k},v'c'\vb{k}'}  A^{S}_{v'c'\vb{k}'}=\Omega_S A^{S}_{vc\vb{k}},
\end{equation}
where $\ket{S}=\sum_{vc\mathbf{k}}A^S_{vc\mathbf{k}}\hat{a}^{\dagger}_{c\mathbf{k}}\hat{a}_{v\mathbf{k}}\ket{0}$~\cite{rohlfing2000electron}. 
$E_{nk}$ are quasiparticle energies obtained from a $G_0W_0$ calculation~\cite{deslippe2012berkeleygw} and $K$ is the electron-hole interaction kernel~\cite{rohlfing2000electron}. Numerical details and convergence checks for linear and nonlinear optical spectra are provided in the Supplementary Information (SI)~\cite{citeSI}.

\begin{figure}
\centering
\includegraphics[width=8.0cm,clip=true]{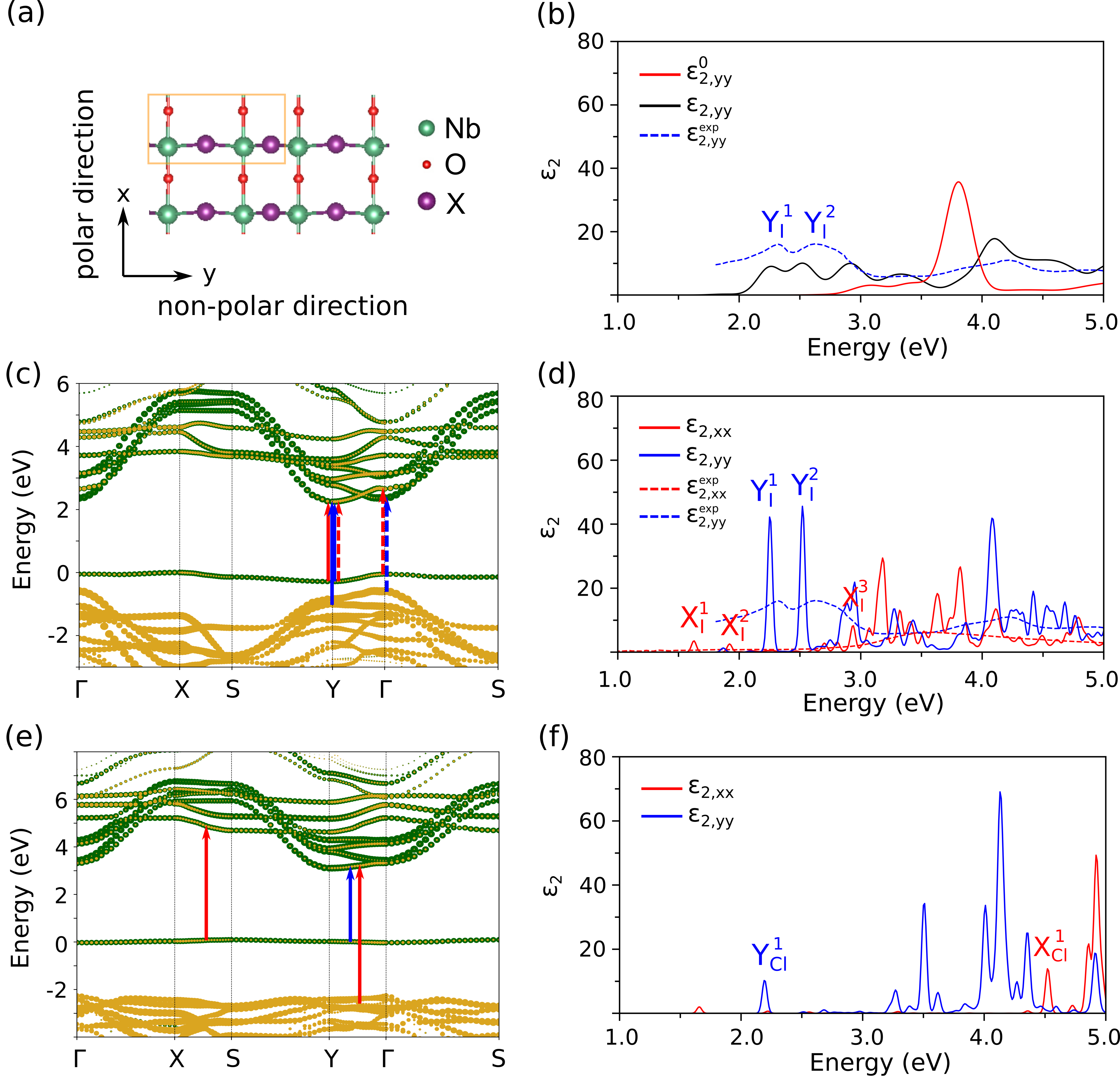}
\caption{\label{fig1}
(a) Atomic structure for ML NbOX$_2$ (X = I, Cl). 
(b) Optical absorption spectra for ML NbOI$_2$ for light polarized in the non-polar direction computed using the $GW$-BSE (black) and $GW$-IPA (red) approaches, compared with experiment~\cite{abdelwahab2022giant} (blue dashed). The broadening value used is $0.10$ eV. 
(c) $GW$ bandstructure for ML NbOI$_2$. 
(d) Optical absorption spectra for ML NbOI$_2$ computed within the $GW$-BSE approach, with a broadening of $0.02$ eV. Red: polar direction, Blue: non-polar direction.
(e,f) Similar to (c-d) but for ML NbOCl$_2$. Experimental data for NbOI$_2$ nanosheets are also shown in (d). 
The green and yellow colors in the band structure denote projections onto Nb and X (X = I, Cl), respectively. 
Selected excitonic resonances are labelled in the absorption spectra. 
Note that X$^3_{\text{I}}$ refers to a group of excitons with very similar energies. 
Dashed blue and solid blue arrows in (c) illustrate the predominant optical transitions contributing to the excitons labelled, respectively, 
by Y$^1_{\text{I}}$ and Y$^2_{\text{I}}$ in (d), and solid red and dashed red arrows illustrate those for excitons X$^1_{\text{I}}$ and X$^2_{\text{I}}$, respectively. 
Likewise, red and blue arrows correspond to X$^1_{\text{Cl}}$ and Y$^1_{\text{Cl}}$ in (f). 
}
\end{figure}


The atomic structure of monolayer (ML) NbOX$_2$ (X=I, Cl) is shown in Figure \ref{fig1}a. Nb atoms are displaced off-center along the $x$ (polar)-direction ~\cite{wu2022data,jia2019niobium}, resulting in three non-vanishing components $\chi^{(2)}_{xxx}$, $\chi^{(2)}_{xyy}$, $\chi^{(2)}_{yxy}$, which can lead to Type 0, I and II SPDC processes respectively~\cite{pan2012multiphoton}. The band structures of ML NbOI$_2$ and NbOCl$_2$ (Figure ~\ref{fig1}c and ~\ref{fig1}e) are computed using many-body perturbation theory within the $GW$ approximation. 
The valence bands in NbOCl$_2$ are deeper than in NbOI$_2$, reflecting the greater ionicity of the Nb-Cl bonds. 
 
Figure \ref{fig1}b shows the computed linear optical response $\varepsilon_{2,yy}$ for NbOI$_2$ compared to experiment~\cite{abdelwahab2022giant} (with a broadening chosen to be similar to the experiment at low energies). There is good agreement to within 0.1 eV between the $GW$-BSE (black) and experimental (blue) peak positions, in particular for Y$^1_{\text{I}}$ and Y$^2_{\text{I}}$. Differences between $GW$-BSE and experiment at higher energies may result from energy-dependent broadening effects~\cite{qiu2013optical}.  In contrast, in the $GW$-IPA result (red), the two peaks labeled by Y$^1_{\text{I}}$ and Y$^2_{\text{I}}$ are clearly absent. Instead, a large peak arises close to $4.0$ eV in the IPA, due to band nesting between two relatively flat Nb-bands; this band nesting effect is completely removed by destructive interference among the dipole matrix elements when electron-hole interactions are taken into account. Figure ~\ref{fig1}d and ~\ref{fig1}f present more $GW$-BSE results for the linear optical response, and compared to the $GW$-IPA results (Figure S12), we see that there is a significant difference in spectral shape, and the BSE response has excitations at lower energies, as we expect from attractive electron-hole interactions. 
\begin{figure*}
\centering
\includegraphics[width=15.0cm,clip=true]{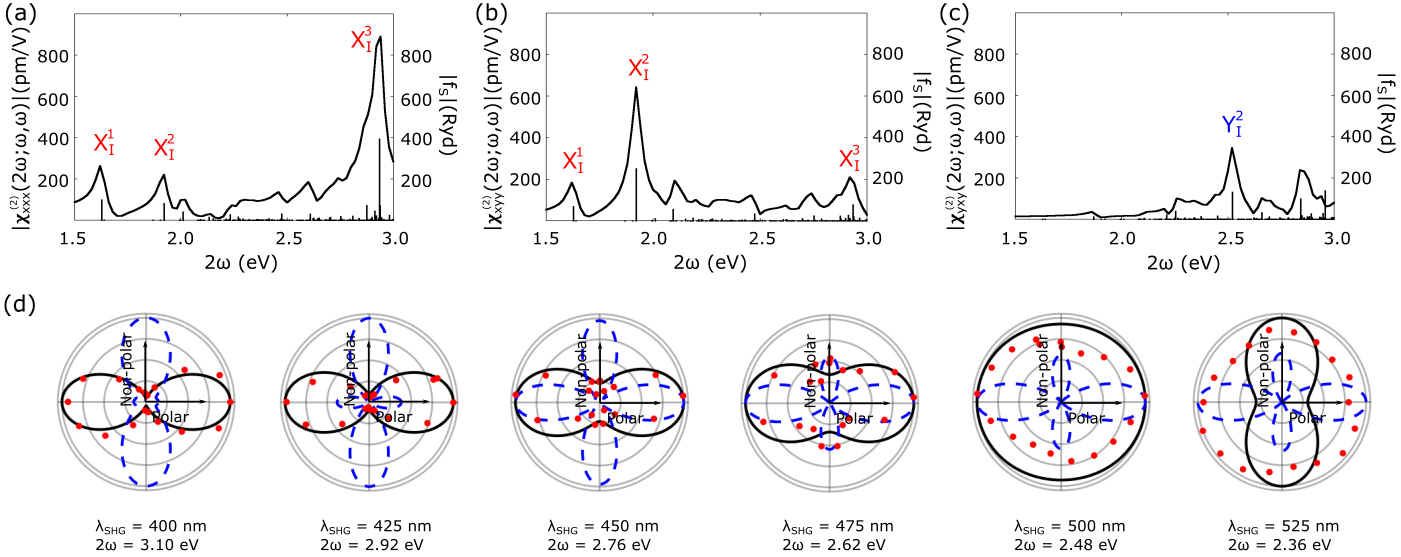}
	\caption{Degenerate SPDC and SHG in NbOI$_2$.
        (a, b, c) Absolute value of (a) $\chi^{(2)}_{xxx}$, (b) $\chi^{(2)}_{xyy}$, (c) $\chi^{(2)}_{yxy}$ (left axis) corresponding to degenerate SPDC or SHG in ML NbOI$_2$. 
	The $2\omega$-resonance strengths $|f^{2\omega}_S|$ (black vertical lines) are plotted against $\Omega_S$ (right axis) (see text). 
	Selected peaks are labelled by the associated excitonic resonances. 
	The lowest $\omega$-resonances are $>3.0$eV as the first excitation in $\varepsilon_2^{xx}$ is $>1.5$ eV. 
	(d) Total SHG intensity plotted as a function of the polarization angle of the incoming light for NbOI$_2$. Solid black lines: $GW$-BSE, Dashed blue lines: $GW$-IPA, Red dots: Experiment~\cite{abdelwahab2022giant}. The absolute values of the computed SHG intensities have been renormalized so that the shape of the polar plots can be seen clearly on the same scale. 
	}
 \label{fig3}
\end{figure*}

\begin{figure*}
\centering
\includegraphics[width=15.0cm,clip=true]{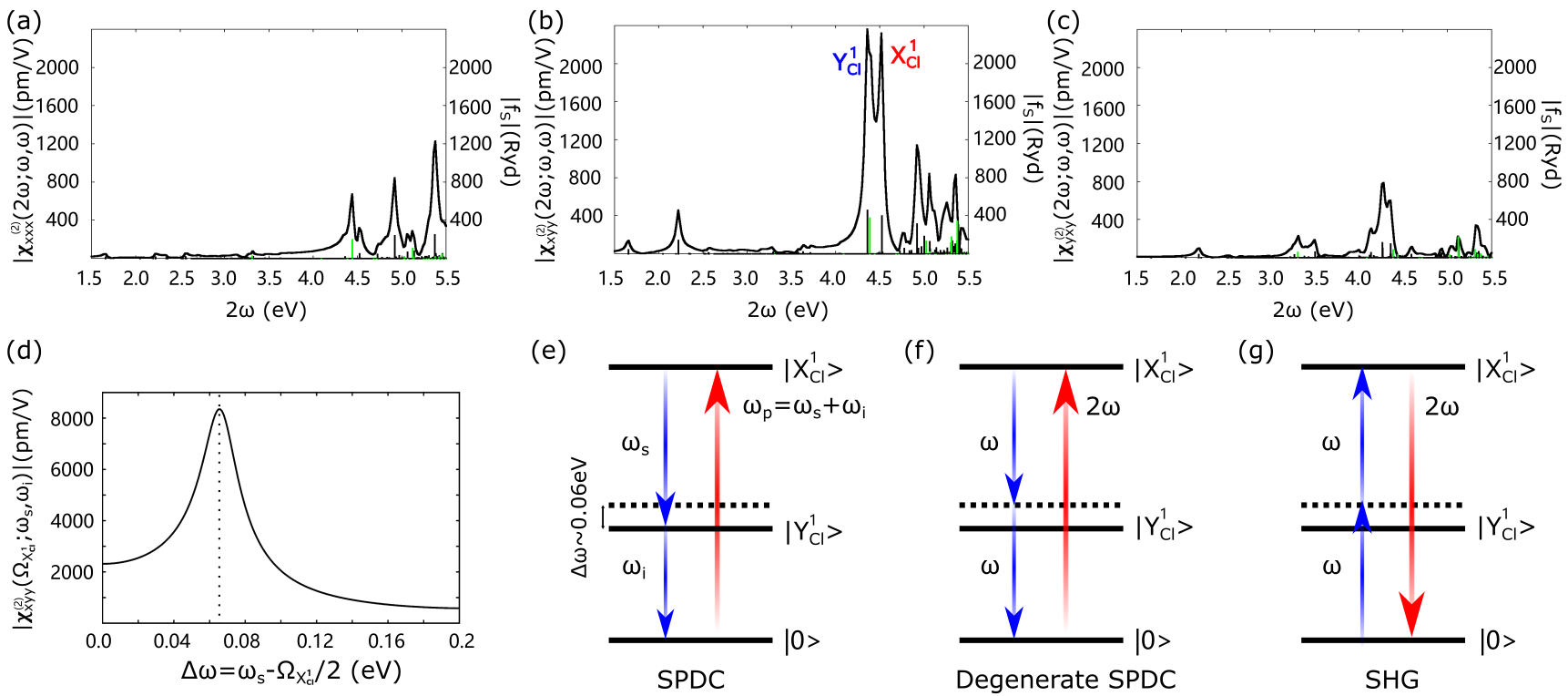}
	\caption{SPDC and SHG in ML NbOCl$_2$.  (a, b, c) Absolute value of (a) $\chi^{(2)}_{xxx}$, (b) $\chi^{(2)}_{xyy}$, (c) $\chi^{(2)}_{yxy}$ (left axis) corresponding to degenerate SPDC or SHG. \textcolor{black}{The $2\omega$-resonance strengths $|f^{2\omega}_S|$ (black vertical lines) are plotted against $\Omega_S$, and the $\omega$-resonance strengths $|f^{\omega}_{S'}|$ (green vertical lines) are plotted against $2\Omega_{S'}$ (right axis). Selected} peaks are labelled by the associated excitonic resonances. (d) Absolute value of $\chi^{(2)}_{xyy}(\omega_p;\omega_s,\omega_i)$ with $\omega_p=\Omega_{\text{X}^1_{\text{Cl}}}$, plotted against $\Delta\omega = \omega_s-\Omega_{\text{X}^1_{\text{Cl}}}/2$, computed using $GW$-BSE. (e)  Schematic figure for non-degenerate SPDC process with the exact double-resonance condition. (f-g) Schematic figures for the degenerate SPDC (f) and SHG (g) processes corresponding to the double-resonance peak in (b). 
	}
 \label{fig4}
\end{figure*}

Next we look at the magnitude of the calculated second order susceptibilities in Figure \ref{fig3} and Figure \ref{fig4}, for NbOI$_2$ and NbOCl$_2$, respectively. 
Compared to the $\chi^{(2)}$ computed within the IPA \textcolor{black}{(see Figure S13)}, we see that excitonic effects enhance the nonlinear response by orders of magnitude and significantly alter its frequency dependence. 
This is in contrast to previous studies on bulk zinc-blende systems, where excitonic effects did not have such drastic effects on the SHG spectra~\cite{2005PRB,chang2001excitonic,riefer2017solving}. 
To investigate whether these giant excitonic effects on $\chi^{(2)}$ can be observed experimentally, we focus on the inverse of the degenerate SPDC process - SHG. 
Figure \ref{fig3}d shows the frequency-dependent polar plots of the total SHG intensity against the polarization angle of the incoming light for NbOI$_2$ (Figure \ref{fig3}d) ~\cite{citeSI}. 
The $GW$-BSE results (black) agree well with experiment~\cite{abdelwahab2022giant} (red dots), except for the plot for $2\omega=2.36$ eV. 
The $GW$-IPA results (dashed blue) deviate significantly from experiment across the energy range shown. 
We emphasize that the better agreement between BSE and experiment is robust to small energy shifts (see Figure S7 and discussion in the SI~\cite{citeSI}). 

The shape of the polar plot is determined by the relative magnitudes and phases of different $\chi^{(2)}$ components (see Figure S7a). 
For this energy range, the contribution from $|\chi^{(2)}_{yxy}|$ is small in comparison to those from $|\chi^{(2)}_{xxx}|$ or $|\chi^{(2)}_{xyy}|$, 
due to absorption of the outgoing $y$-polarized light~\cite{citeSI}. 
For $2\omega\sim2.7-2.8$ eV, $|\chi^{(2)}_{xyy}|>|\chi^{(2)}_{xxx}|$ in the IPA (Figure S13), but $|\chi^{(2)}_{xxx}|>|\chi^{(2)}_{xyy}|$ for BSE. 
As we discuss below, the peaks in $\chi^{(2)}$ in Figure \ref{fig3} are attributed to excitonic resonances with $2\omega \sim \Omega_S$. 
We note that the IPA transitions in $\sim2.7-2.8$ eV are the dominant transitions in the $\text{X}^2_{\text{I}}$ exciton at $\sim1.9$ eV. 
In both IPA and BSE, $|\chi^{(2)}_{xyy}|>|\chi^{(2)}_{xxx}|$ for the $2\omega$-resonance corresponding to this transition; 
in this case, the discrepancy between IPA and experiment is due to the energy shift from the exciton binding energy. 
For $2\omega\sim2.4-2.5$ eV, we see, interestingly, experimental evidence that the relative phases of the different $\chi^{(2)}$ components matter - in both the experiment and BSE plots, $\chi^{(2)}_{xyy}$ and $\chi^{(2)}_{xxx}$ have similar phases, in contrast to the IPA (Figure S14). 
This comparison illustrates the fundamental excitonic origins of the experimentally-observed SHG~\cite{abdelwahab2022giant}.

To provide deeper insights into the contribution of excitonic resonances to $\chi^{(2)}$, we define $2\omega$-resonance and $\omega$-resonance strengths, $f_{S}^{2\omega}$ and $f_{S}^{\omega}$. 
These quantities describe the contribution to $\chi^{(2)}$ arising from resonances of the incoming/outgoing photon energies with the many-body excitation energies. 
Keeping the terms in Eq.~\ref{chi} which contribute the most at positive $\omega$, we have
\textcolor{black}{
\begin{equation}
\begin{split}
        \chi_{\alpha\beta\gamma}^{(2)}(2\omega;\omega,\omega) \approx \frac{-ie^3}{m^3\Tilde{\omega}^3}\sum_{SS'}
 \frac{\alpha_{0S} \{\beta_{SS'} \gamma_{S'0}\}}{(\Tilde{\omega}-\Omega_{S'})(2\Tilde{\omega}-\Omega_S)} \\
 \approx \frac{e^3}{m^3}[\sum_{S}
 \frac{2f_{S}^{2\omega}}{2\Tilde{\omega}-\Omega_S} 
 +\sum_{S'}\frac{f_{S'}^{\omega}}{\Tilde{\omega}-\Omega_{S'}}]
\end{split}
\end{equation} 
}
\textcolor{black}{
where $ \{\beta_{SS'} \gamma_{S'0}\}=(\beta_{SS'} \gamma_{S'0}+\gamma_{SS'} \beta_{S'0})/2$, and 
\begin{equation}
    f_{S}^{2\omega} = \frac{8}{\Omega_S^3}\sum_{S'}
 \frac{-i\alpha_{0S} \{\beta_{SS'} \gamma_{S'0}\}}{\Omega_S-2\Omega_{S'}},  
\label{f2}
\end{equation}
}
\textcolor{black}{
\begin{equation}
    f_{S'}^{\omega} = -\frac{1}{\Omega_{S'}^3} \sum_{S}
 \frac{-i\alpha_{0S}\{ \beta_{SS'} \gamma_{S'0}\}}{\Omega_S-2\Omega_{S'}}. 
\label{f}
\end{equation}
}

Because the energy axis for $\chi^{(2)}$ corresponds to the outgoing energy $2\omega$, in order to illustrate the contributions to peaks in $\chi^{(2)}$, 
we plot $|f_S^{2\omega}|$ against $\Omega_S$ and $|f_{S'}^{\omega}|$ against $2\Omega_{S'}$ (the incoming photon energy is resonant with $\Omega_{S'}$) (Fig.~\ref{fig3}(a-c) and Fig.~\ref{fig4}(a-c)). 
We label selected peaks in $\lvert \chi^{(2)}(2\omega;\omega,\omega) \rvert$ with the excitons that predominantly contribute to the peaks through the resonance effect and large resonance strengths $f$. 
For the energy range shown here, most of the peaks in $\chi^{(2)}$ arise from $2\omega$-resonances (black), except for a few $\omega$-resonances in NbOCl$_2$ (green; Fig.~\ref{fig4}). 
The resonance strengths for each exciton $S$ arise from the collective effect of interexciton transitions $\mel{S}{\vb{P}}{S'}$, $S'\neq S$, where $S'$ involves both low and high energy excitations~\cite{citeSI} (see Figure S15). Terms involving intraexciton transitions ($\mel{S}{\bf{P}}{S}$) can only contribute to $\chi^{(2)}$ in systems with broken time-reversal symmetry~\cite{citeSI}. For a fixed exciton $S$ (e.g.  X$_\text{I}^2$ or  X$_\text{I}^3$), the observed differences in $|f_S^{2\omega}|$ for $\chi^{(2)}_{xxx}$ and $\chi^{(2)}_{xyy}$ can be primarily attributed to differences in $\mel{S}{P^x}{S'}$ and $\mel{S}{P^y}{S'}$ across the range of excitons $S'$ (see e.g. Figure S16). In particular, we also note that $|f_S|$ can be very large even when the oscillator strength for $S$ is small, leading to large nonlinear optical responses.

The order-of-magnitude larger $\lvert \chi^{(2)}\rvert$ values in BSE compared to the IPA can be attributed to the correspondingly larger values of $|f_S^{2\omega}|$ and $|f_{S'}^{\omega}|$ (Fig.~\ref{fig3}-\ref{fig4} vs Fig. S13). Exciton states are a linear combination of interband transitions at different $k$-points. This opens up more interexciton transition channels as can be seen from the BSE and corresponding IPA expressions for $\mel{S}{\vb{P}}{S'}$:
\begin{equation}
\begin{split}
\mel{S}{\vb{P}}{S'} = &\sum_{vcc'\vb{k}} A_{vc\vb{k}}^{S*}A_{vc'\vb{k}}^{S'}\vb{p}_{cc'\vb{k}} - 
 \sum_{v'vck} A_{vc\vb{k}}^{S*}A_{v'c\vb{k}}^{S'}\vb{p}_{v'v\vb{k}} \\
        \xrightarrow[\ket{S'}=\ket{v_2c_2\vb{k}_2}]{\ket{S}=\ket{v_1c_1\vb{k}_1}}&\quad \vb{{p}}_{c_1c_2\vb{k}_2} \delta_{\vb{k}_1\vb{k}_2} \delta_{v_1v_2} -\vb{\vb{p}}_{v_2v_1\vb{k}_1}\delta_{\vb{k}_1\vb{k}_2} \delta_{c_1c_2}.
\end{split}
\end{equation}
When the exciton is reduced to a momentum-conserving interband transition between Bloch states in the IPA, many $\mel{S}{\vb{P}}{S'}$ elements become zero due to the collapse of the exciton wavefunction to a single point in reciprocal space~\cite{lai2024bulk}. 
 Furthermore, in the BSE case, peaks in $\chi^{(2)}$ are observed to correspond to sharp spikes in $|f_S|$, i.e. a few excitons $S$ have particularly large resonance strengths; this is not the case in the IPA (Figure S13), where the variation in $|f_S|$ is smaller. We expect these excitonic effects, leading to enhanced SPDC/SHG, and stronger and sharper resonance contributions, to be larger in 2D and quasi-2D layered materials compared to bulk 3D systems due to the reduced electronic screening in 2D materials.



Comparing NbOI$_2$ and NbOCl$_2$, we recall that in NbOCl$_2$, the valence bands are deeper, and consequently, there are fewer excitations in the lower energy range. We therefore show $\lvert \chi^{(2)} \rvert$ at a higher energy range in Figure \ref{fig4}.
We observe a giant double excitonic resonance feature at $\sim 4.5$ eV for $\chi^{(2)}_{xyy}$ in NbOCl$_2$ (Figure \ref{fig4}b), 
as the energy of the Y$^1_{\text{Cl}}$ exciton is approximately half that of X$^1_{\text{Cl}}$; both $\omega$ and $2\omega$ are nearly resonant with one of the exciton energies (Figure \ref{fig4}b, f-g). We generalize our discussion of the double resonance feature to the case of non-degenerate SPDC, the generation of entangled photons with different frequencies, with $\omega_s = \frac{\omega_p}{2}+\Delta\omega$ and $\omega_i = \frac{\omega_p}{2}-\Delta\omega$. 
Figure \ref{fig4}d shows the predicted $\lvert \chi^{(2)}_{xyy} \rvert$ for an incoming photon resonant with X$^1_{\text{Cl}}$, as a function of $\Delta\omega$; it is clear that $\lvert \chi^2_{xyy} \rvert$ increases significantly as the exact \textcolor{black}{double-exciton} resonance condition (for both $\omega_p$ and $\omega_s$) (Figure \ref{fig4}e) is approached. 

Our work provides convincing evidence of the fundamental excitonic origins of nonlinear optical responses in layered 2D materials, and calls for future experimental studies to realise potentially higher yields of entangled photons using the double exciton resonance condition. 
This paves the way for 2D sources of correlated photons in next-generation quantum photonics devices.

\textit{Note added.} After this manuscript was submitted, a paper on excitonic effects in SHG for 2D materials appeared in the arXiv~\cite{ruan2023excitonic}.

\section{Acknowledgements}
This work is supported by NUS and the National Research Foundation (NRF), Singapore, under the NRF medium-sized centre programme.
Calculations were performed at the National Supercomputing Centre, Singapore, on thecluster in the Centre for Advanced 2D Materials and on Fugaku provided by RIKEN through the HPCI System Research Project (Project ID: hp220152). 
The majority of the work by F.X. was performed in NUS.
We thank I. Abdelwahab for discussions on the experimental data cited here.
\bibliographystyle{apsrev4-1}

\begin{thebibliography}{1}
\bibitem{dorfman2016nonlinear} K. Dorfman, F. Schlawin, and S. Mukamel, {\em Rev. Mod. Phys.} \textbf{88}, 045008 (2016).
\bibitem{reverseCouteau} C. Couteau, {\em Contemp. Phys.} \textbf{59}, 291-304 (2018).
\bibitem{pan2012multiphoton}J.-W. Pan, Z.-B. Chen, C.-Y. Lu, H. Weinfurter, A. Zeilinger \& M. Żukowski, {\em Rev. Mod. Phys.} \textbf{84}, 777 (2012).
\bibitem{guo2023ultrathin} Q. Guo, X. Qi, L. Zhang, M. Gao, S. Hu, W. Zhou, W. Zang, X. Zhao, J. Wang, B. Yan, and et. al. {\em Nature} \textbf{613}, 53-59 (2023).
\bibitem{ResNanoPh2021} Q. Leng, H. Su, J. Liu, L. Zhou, K. Qin, Q. Wang, J. Fu, S. Wu, and X. Zhang, {\em Nanophotonics} \textbf{10}, 1871-1877 (2021).
\bibitem{ResOpt2021}T. Santiago-Cruz, V. Sultanov, H. Zhang, L. Krivitsky, and M. Chekhova, {\em Optics Letters} \textbf{46}, 653-656 (2021).
\bibitem{ResACSnano2022} J. Shi, Z. Lin, Z. Zhu, J. Zhou, G. Xu, and Q. Xu, {\em ACS Nano} \textbf{16} 15862-15872 (2022).
\bibitem{PRLthin} C. Okoth, A. Cavanna, T. Santiago-Cruz, and M. V. Chekhova, {\em Phys. Rev. Lett.} \textbf{123} 263602 (2019).
\bibitem{levine1990one} Z. Levine, {\em Phys. Rev. B} \textbf{42}, 3567 (1990).
\bibitem{sipe1993nonlinear}J. Sipe, and E. Ghahramani, {\em Phys. Rev. B} \textbf{48}, 11705 (1993).
\bibitem{luppi2010ab}E. Luppi, H. Hubener, and V. Veniard, {\em Phys. Rev. B} \textbf{82}, 235201 (2010).
\bibitem{young2012first} S. Young, and A. Rappe, {\em Phys. Rev. Lett.} \textbf{109}, 116601 (2012).

\bibitem{rangel2017large} T. Rangel, B. Fregoso, B. Mendoza, T. Morimoto, J. Moore, and J. Neaton, {\em Phys. Rev. Lett.} \textbf{119}, 067402 (2017).
\bibitem{Thygesen2021ACSnano}A. Taghizadeh, K. S. Thygesen, and T. G. Pederson, {\em ACS Nano} \textbf{15}, 7155-7167 (2021).
\bibitem{wang2018colloquium}G. Wang, A. Chernikov, M. Glazov, T. Heinz, X. Marie, T. Amand, and B. Urbaszek, {\em Rev. Mod. Phys.} \textbf{90}, 021001 (2018).
\textcolor{black}{
\bibitem{attaccalite2011real} C. Attaccalite, M. Gr\"{u}ning, and A. Marini, {\em Phys. Rev. B} \textbf{84}, 245110 (2011).
\bibitem{attaccalite2013nonlinear} C. Attaccalite and M. Gr\"{u}ning, {\em Phys. Rev. B} \textbf{88}, 235113 (2013).
\bibitem{kolos2021giant} M. Kolos, L. Cigarini, R. Verma, F. Karlick\'{y} and S. Bhattacharya, {\em J. Phys. Chem. C} \textbf{125}, 12738 (2021).
\bibitem{chan2021giant} Y.-H. Chan, D. Y. Qiu, F.H. da Jornada and S.G. Louie, {\em PNAS} \textbf{118}, e1906938118 (2021).
}
\bibitem{2005PRB} R. Leitsmann, W. Schmidt, P. Hahn, and F. Bechstedt, {\em Phys. Rev. B} \textbf{71}, 195209 (2005).
\bibitem{chang2001excitonic} E. K. Chang, E. L. Shirley, and Z. H. Levine, {\em Phys. Rev. B} \textbf{65}, 035205 (2001).
\bibitem{riefer2017solving} A. Riefer, and W. G. Schmidt, {\em Phys. Rev. B} \textbf{96}, 235206 (2017).


\bibitem{abdelwahab2022giant} I. Abdelwahab, B. Tilmann, Y. Wu, D. Giovanni, I. Verzhbitskiy, M. Zhu, R. Bert\'{e}, F. Xuan, L. Menezes, G. Eda, and et. al., {\em Nat. Photonics} \textbf{16}, 644-650 (2022).
\bibitem{abdelwahab2023highly} I. Abdelwahab, B. Tilmann, X. Zhao, I. Verzhbitskiy, R. Bert\'{e}, G. Eda, W. Wilson, G. Grinblat, and et al, {\em Adv. Opt. Mater.} 2202833 (2023).
\bibitem{AdvMaterNbOI2} Y. Fang, F. Wang, R. Wang, T. Zhai and F. Huang, {\em Adv. Mater.} \textbf{33} 2101505 (2021).
\bibitem{Mandel1985} C. Hong, and L. Mandel, {\em Phys. Rev. A} \textbf{31}, 2409 (1985).
\bibitem{Giovanni2021} H. Rostami, M. I. Katsnelson, G. Vignale, and M. Polini, {\em Ann. Phys. } \textbf{431}, 168523 (2021).
\bibitem{BOYD2008135}R. Boyd, {\em Nonlinear Optics (Third Edition)} (Academic Press, Burlington, 2008). pp. 135-206
\bibitem{ahn2020low}J. Ahn, G. Guo, and N. Nagaosa, {\em Phys. Rev. X}. \textbf{10}, 041041 (2020)

\bibitem{rohlfing2000electron}M. Rohlfing, and S. Louie, {\em Phys. Rev. B} \textbf{62}, 4927 (2000).
\bibitem{deslippe2012berkeleygw} J. Deslippe, G. Samsonidze, D. Strubbe, M. Jain, M. Cohen, and S. Louie, {\em Comput. Phys. Commun.} \textbf{183}, 1269-1289 (2012).
\bibitem{citeSI}See Supplemental Material at [url], which includes Refs. [33–41], for computational details and numerical convergence tests.
\bibitem{Grimme} S. Grimme et al \textit{J. Comput. Chem.} \textbf{27}, 1787 (2006)
\bibitem{QE} P. Giannozzi, \textit{J. phys. Condens. Matter} \textbf{21} 395502 (2009)%
\bibitem{ONCV} D.R. Hammann \textit{Phys. Rev. B} \textbf{88}, 085117 (2013)
\bibitem{slab} S. Ismail-Beigi \textit{Phys. Rev. B} \textbf{73}, 233103 (2006) 
\bibitem{1986} M.S. Hybertsen and S.G. Louie, \textit{Phys. Rev. B} \textbf{34}, 5390 (1986)
\bibitem{Felipe} H. Felipe et al., \textit{Phys. Rev. B} \textbf{95}, 035109 (2017)
\bibitem{shen1984principles} Y.-R. Shen \textit{Principles of nonlinear optics} Wiley-Interscience, New York, NY, USA, (1984)%
\bibitem{Cohen-Tannoudji} C. Cohen-Tannoudji, \textit{Quantum Mechanics} 2, 626 (1986)%
\bibitem{Chen} H.-Y. Chen \textit{Phys. Rev. Lett.} \textbf{125}, 107401 (2020)%
\bibitem{wu2022data} Y. Wu, I. Abdelwahab, K. Kwon, I. Verzhbitskiy, L. Wang et al., {\em Nat. Commun.} \textbf{13}, 1884 (2022).
\bibitem{jia2019niobium}Y. Jia, M. Zhao, G. Gou, X. Zeng, and J. Li, {\em Nanoscale Horiz.} \textbf{4}, 1113-1123 (2019).

\bibitem{qiu2013optical} D. Qiu, H. Felipe, and S. Louie, {\em Phys. Rev. Lett.} \textbf{111}, 216805 (2013).
\bibitem{lai2024bulk} M. Lai, F. Xuan, and S.Y. Quek, {\em arXiv} 2402.02002 (2024).
\bibitem{ruan2023excitonic} J. Ruan, Y.H. Chan, and S. Louie, {\em arXiv} 2310.09674 (2024).

\end{thebibliography}

\end{document}